\documentclass[final,3p,times,twocolumn]{elsarticle2}
\usepackage{graphicx}
\usepackage{nicefrac}
\usepackage{amssymb}
\usepackage[squaren]{SIunits}

\journal{ArXiv.org/cond-mat}

\begin{document}

\begin{frontmatter}

\title{Comment on: ``Revealing common artifacts due to ferromagnetic inclusions in highly oriented pyrolytic graphite'', {\sl by M.~Sepioni, R.R.~Nair, I.-Ling Tsai, A.K.~Geim and I.V.~Grigorieva, EPL 97 (2012) 47001}}

\author[ul]{D. Spemann\corref{corr1}}
\cortext[corr1]{D. Spemann}
\ead{spemann@uni-leipzig.de}

\author[ul]{M.~Rothermel}
\author[ul]{P.~Esquinazi}

\author[cmam]{M.~Ramos}

\author[camp]{Y.~Kopelevich}

\author[stan]{H.~Ohldag}

\address[ul]{Universit\"{a}t Leipzig, Institute for Experimental Physics II, Linn\'estr. 5, 04103 Leipzig, Germany}

\address[cmam]{CMAM and Instituto de Ciencia de Materiales ``Nicol\'{a}s Cabrera'', Universidad Aut\'{o}noma de Madrid, Cantoblanco, Madrid, Spain}

\address[camp]{Instituto de F\'{\i}isica ``Gleb Wataghin" DFA, Universidade Estadual de Campinas-UNICAMP,
Cidade Universitaria Zeferino Vaz, Bairro,\\ Bar\~{a}o Geraldo 13083-859, Brasil}

\address[stan]{Stanford Synchrotron Radiation Lightsource, Stanford University, Menlo Park, CA 94025, USA}

\end{frontmatter}

Herewith we would like to comment on the recently reported study on ``common artifacts'' due to impurities published by Sepioni et al.$\!$ \cite{1} related to the magnetic order measured in highly oriented pyrolytic graphite (HOPG) samples. Apparently the authors in \cite{1} are not aware of the proper use of the Particle Induced X-ray Emission (PIXE) technique and its ability for elemental imaging with lateral resolution of $\approx\!1$\,\micro m. As it is stated in previous publications \cite{2,3} PIXE is typically carried out using 2000\,keV protons (not 200\,keV as stated in Ref.~\cite{1}), which penetrate $38$\,\micro m deep into graphite and excite X-rays efficiently up to a depth of $\approx\!20$\,\micro m. The absorption of the {\sl K}-X-rays of e.g.$\!$ Ti, V, Cr, Fe and Ni on their way back to the detector can safely be neglected in graphite making the PIXE technique a truly bulk-sensitive technique with excellent sensitivity well below 1\,\micro g/g. PIXE is able to measure in-situ the total amount of Fe in bulk graphite contrary to EDX, which needs to analyze individual particles of contamination due to its comparably poor minimum detection limits. Therefore, PIXE allows to determine directly the metallic content as has also been proved in a comparison with Neutron Activation Analysis (NAA) on ZYB graphite where both techniques gave a Fe concentration of 0.17\,\micro g/g \cite{4}.

Metallic particles reported by the authors in \cite{1} are well-known in ZY graphite and have already been reported as well \cite{4,5}. We would like to point out that the HOPG samples from NT-MDT used in Ref.~\cite{1} are known to contain much more potentially magnetic contamination than, e.g., HOPG from Advanced Ceramics, which has been used in most of the published work by the authors of this comment \cite{2,3,4,5,7,9}. Certainly, there is always the possibility of contamination in any sample that is why each sample has to be analyzed individually using PIXE or any other method with similar capabilities. Apart from that, there are other, element specific microscopic techniques like X-ray Magnetic Circular Dichroism (XMCD) that independently proved the intrinsic, impurity independent origin of ferromagnetism on pure graphite samples before and after proton irradiation \cite{5}  as well as in graphitic thin films \cite{6}.

Furthermore, it is important to emphasize that whatever the ultimate origin of the ferromagnetic signals found by SQUID magnetometry, also in pristine, unirradiated ZY graphite samples, one can clearly rule out that ``...the contamination could be the reason for the reported ferromagnetism'', as the authors claimed in \cite{1}. In most of the alluded works \cite{3,7,8,9}, and in some other publications, the net magnetization difference between the same sample after and before ion irradiation is plotted, or else both measurements are shown together. Also, subsequent irradiation and further annealing or ageing of a given sample can modify \cite{7,8} the defect-induced magnetization, as expected. Therefore, the reported ferromagnetic and paramagnetic contributions after ion beam irradiation of ZY graphite samples cannot be due to any kind of ``intrinsic'' magnetic impurities, which would remain constant all along, but they are induced by the ion beam irradiation process and related to the generated defects. We therefore recommend the authors in \cite{1} to read the nowadays broad literature on the defect-induced magnetism (DIM) in pure carbon-based and other materials more carefully.

\section*{References}
\bibliographystyle{elsarticle-num}
\bibliography{Comment_EPL97_Sepioni}

\end{document}